\documentclass[aps,prx,twocolumn,floatfix,longbibliography,superscriptaddress]{revtex4-2}

\usepackage{amsmath}
\usepackage{amssymb}
\usepackage{times}
\usepackage{braket}
\usepackage[pdftex]{graphicx}
\usepackage{color}
\usepackage{blindtext}
\usepackage{nicefrac}
\usepackage[colorlinks=true, citecolor=blue, urlcolor=blue, linkcolor=blue]{hyperref}

\renewcommand{\vec}[1]{\mathbf{#1}}

\newcommand{\change}[1]{\textcolor{black}{#1}}
\newcommand{\changeb}[1]{\textcolor{black}{#1}}
\renewcommand{\ket}[1]{\lvert#1\rangle} 
\newcommand{\braopket}[3]{\langle #1 | #2 | #3\rangle} 

\begin{document}

\title{Chirality-induced selectivity of angular momentum by orbital Edelstein effect in carbon nanotubes}

\author{B{\"o}rge G{\"o}bel}
\email[Correspondence email address: ]{boerge.goebel@physik.uni-halle.de}
\affiliation{Institut f\"ur Physik, Martin-Luther-Universit\"at Halle-Wittenberg, D-06099 Halle (Saale), Germany}

\author{Ingrid Mertig}
\affiliation{Institut f\"ur Physik, Martin-Luther-Universit\"at Halle-Wittenberg, D-06099 Halle (Saale), Germany}

\author{Samir Lounis}
\affiliation{Institut f\"ur Physik, Martin-Luther-Universit\"at Halle-Wittenberg, D-06099 Halle (Saale), Germany}

\date{\today}

\begin{abstract}
Carbon nanotubes (CNTs) are promising materials exhibiting exceptional strength, electrical conductivity, and thermal properties, making them promising for various technologies. Besides achiral configurations with a zigzag or armchair edge, there exist chiral CNTs with a broken inversion symmetry. Here, we demonstrate that chiral CNTs exhibit chirality-induced orbital selectivity (CIOS), which is caused by the orbital Edelstein effect and could be detected as chirality-induced spin selectivity (CISS). We find that the orbital Edelstein susceptibility is an odd function of the chirality angle of the nanotube and is proportional to its radius. For metallic CNTs close to the Fermi level, the orbital Edelstein susceptibility increases quadratically with energy. This makes the CISS and CIOS of metallic chiral nanotubes conveniently tunable by doping or applying a gate voltage, which allows for the generation of spin- and orbital-polarized currents. The possibility of generating large torques makes chiral CNTs interesting candidates for technological applications in spin-orbitronics and quantum computing.  
\end{abstract}

\maketitle


\noindent Carbon nanotubes (CNTs) are an allotrope of carbon and have diameters on the nanometer scale. Since their discovery~\cite{radushkevich1952nanotubes,oberlin1976filamentous,iijima1993single}, the one-dimensional quantum objects have found applications in photonics, electronics, biomedicine, for energy harvesting and storing, water treatment, in displays and many others~\cite{yang2020chirality,de2013carbon,rao2018carbon}. In the physical sciences, CNTs have even been used for logic gates~\cite{huang2001logic}, quantum  technology~\cite{baydin2022carbon} and in devices such as the nano-RAM~\cite{gervasi2019will}, carbon nanotube transistors~\cite{tans1998room,bishop2020fabrication} or microprocessors~\cite{hills2019modern}. 

CNTs can be constructed mathematically by `cutting' a rectangle out of the two-dimensional carbon allotrope graphene and rolling it to a cylinder. The rotation angle $\alpha$ between the rectangle and the underlying graphene lattice influences the resulting physical properties; most famously, some tubes are metallic and others are \change{insulating}. While the two most well-known types of CNTs -- tubes with armchair or zigzag edges -- posses inversion symmetry, all other nanotubes characterized by $0^\circ<|\alpha|<30^\circ$ have a broken inversion symmetry which makes them chiral. Over the recent years, such chirality-pure nanotubes have become more easily available (cf. review~\cite{yang2020chirality}), such as the chiral $(12,6)$ CNTs that were grown with a Co$_7$W$_6$ or Mo$_2$C catalyst with a purity of more than $90\%$~\cite{yang2014chirality,zhang2017arrays}. Furthermore, dispersion by surfactants~\cite{green2011nearly} and other methods~\cite{yang2020chirality} allow for sorting grown tubes based on their chirality.

\begin{figure}[t!]
    \centering
    \includegraphics[width=1\columnwidth]{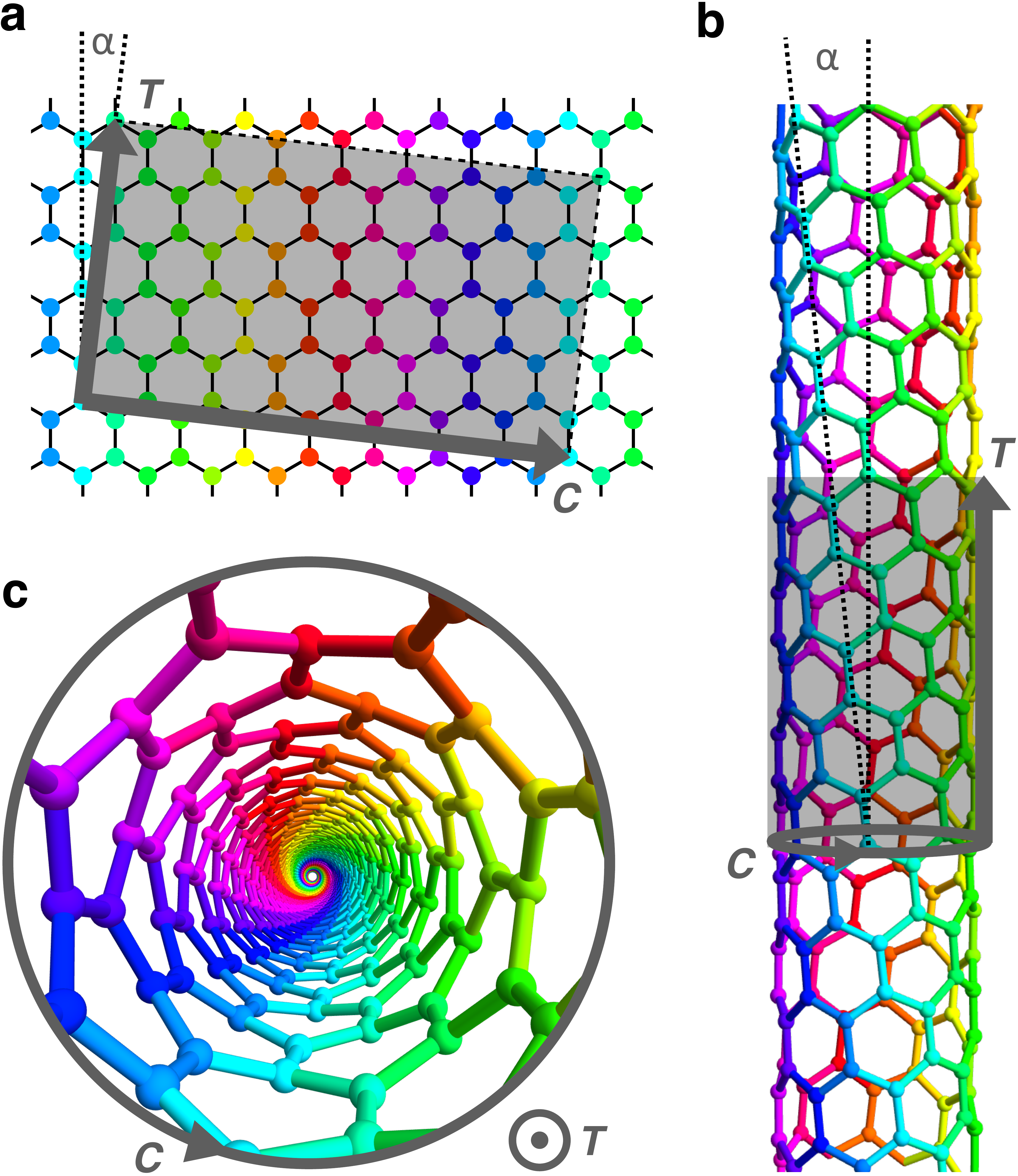}
    \caption{\textbf{Chiral carbon nanotube.} \textbf{a} Geometrical construction of a $(n,m)=(7,1)$ nanotube from a two-dimensional graphene lattice. The color indicates the $x$ coordinate of each atom. The circumferential vector $\vec{C}$ and the translational vector $\vec{T}$ span a rectangle containing the atoms of the unit cells of the corresponding chiral carbon nanotube in \textbf{b}. The chirality angle $\alpha$ is indicated. \textbf{c} Top view of the nanotube.}
    \label{fig:nanotubes}
\end{figure}

A chiral CNT has a handedness and is distinct from its mirror image, similar to a helix that describes the structure of chiral molecules~\cite{ray1999asymmetric}, like the DNA molecule, or periodic crystals like tellurium and selenium~\cite{furukawa2017observation,furukawa2021current,calavalle2022gate}. In these materials, an applied current gives rise to a chirality-induced selectivity of angular momentum. In most of the literature, the spin degree of freedom is discussed; a phenomenon called chirality-induced spin selectivity (CISS)~\cite{ray1999asymmetric,lu2019spin,huang2020magneto,furukawa2017observation,furukawa2021current,calavalle2022gate,inui2020chirality,nabei2020current,alam2015spin,alam2017spin,rahman2018long,rahman2020carrier,rahman2021molecular,hao2021direct,rahman2022chirality,moharana2025chiral}. The structure's chirality acts as a spin filter and determines the orientation of the spin-polarization of an electric current flowing through it. Indeed, CISS has been observed in chiral systems related to CNTs~\cite{alam2015spin,alam2017spin,rahman2018long,rahman2020carrier,rahman2021molecular,hao2021direct,rahman2022chirality} as well, like molecular-functionalized CNTs or chiral heterostructures based on CNTs but the theoretical understanding of the role of the nanotubes' chirality is missing.

\begin{figure*}[t!]
    \centering
    \includegraphics[width=1\textwidth]{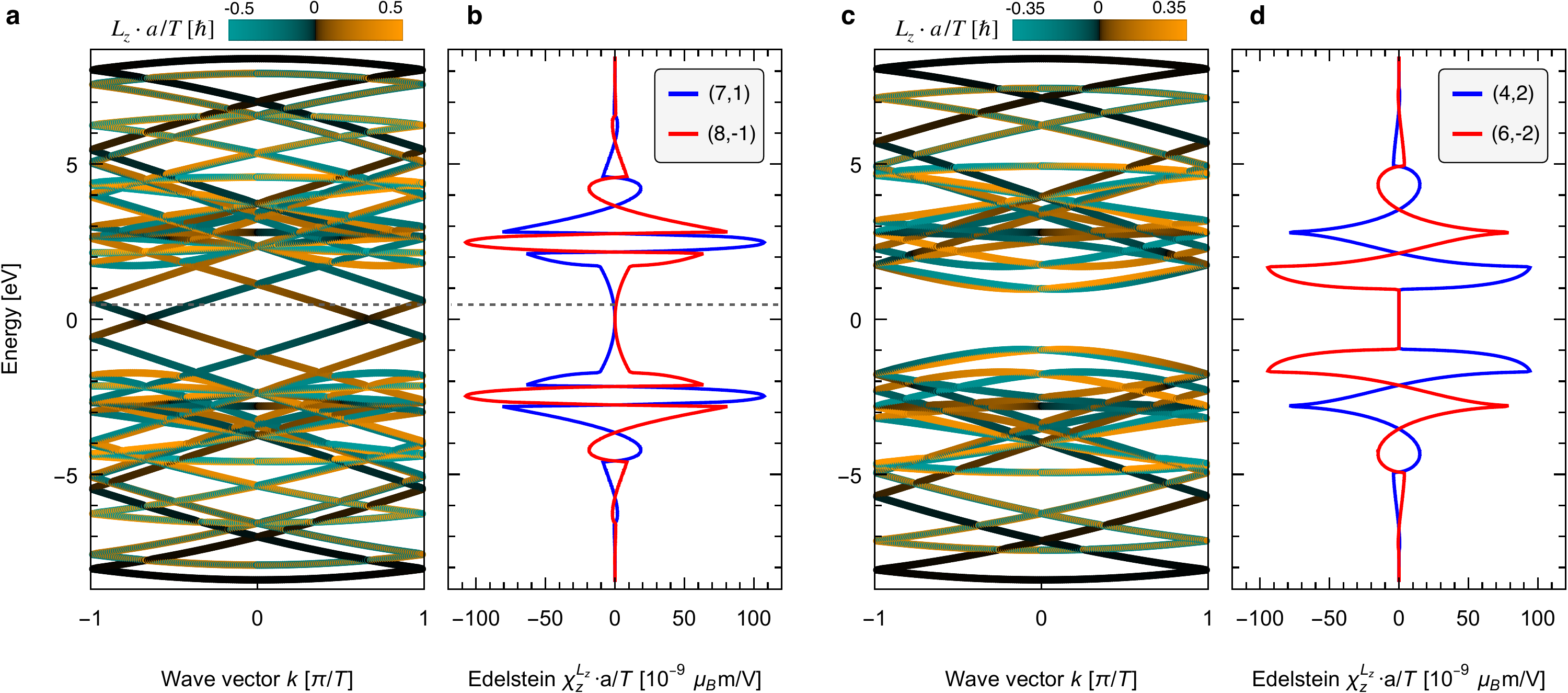}
    \caption{\textbf{Chirality-induced orbital selectivity of a chiral carbon nanotube.} \textbf{a} Band structure of a metallic $(n,m)=(7,1)$ nanotube for which the color indicates the value of the orbital angular momentum $L_{\nu,z}(k)$ normalized to a unit tube length (see legend). \textbf{b} Orbital Edelstein susceptibility $\chi_z^{L_z}\cdot a/T$ as a function of energy for a $(n,m)=(7,1)$ nanotube (blue) and for the corresponding nanotube with opposite chirality $(n,m)=(8,-1)$ (red). The dashed line indicates an energy of $E=0.49\,\mathrm{eV}$ in the range where the orbital Edelstein susceptibility increases quadratically with energy. \textbf{c} Band structure and \textbf{d} orbital Edelstein susceptibility (blue) of an insulating $(n,m)=(4,2)$ nanotube and for the corresponding nanotube with opposite chirality $(n,m)=(6,-2)$ (red). }
    \label{fig:bands_edelstein}
\end{figure*}

Over the past few years, the orbital Hall effect~\cite{zhang2005intrinsic, bernevig2005orbitronics, kontani2008giant, tanaka2008intrinsic, kontani2009giant,go2018intrinsic, pezo2022orbital,salemi2022theory,busch2023orbital,choi2023observation,lyalin2023magneto,busch2024ultrafast,gobel2024OHE,gobel2024topological} and the orbital Edelstein effect~\cite{levitov1985magnetoelectric,yoda2015current,yoda2018orbital, go2017toward,tsirkin2018gyrotropic,salemi2019orbitally,johansson2021spin,liu2021chirality,kim2023optoelectronic,el2023observation,hagiwara2024orbital} have been established as the counterparts to the spin Hall effect~\cite{dyakonov1971current,kato2004observation} and spin Edelstein effect~\cite{aronov1989nuclear,edelstein1990spin}. It has been calculated that the orbital Edelstein effect might explain the large measured CISS: It might mostly be caused by chirality-induced orbital selectivity (CIOS) that is only converted to CISS upon measurement~\cite{yoda2015current,yoda2018orbital,liu2021chirality,kim2023optoelectronic,yen2024controllable,hagiwara2024orbital,gobel2025chirality}. The orbital angular momentum is transformed to spin angular momentum due to the spin-orbit coupling in the electrodes. In a recent study, we have shown that CIOS in a helix structure like in tellurium has a purely geometrical origin and can be explained even classically based on a helical trajectory of a moving electron~\cite{gobel2025chirality}. Orbital angular momentum is generated by a rotational motion and is transported along the helix. The effect is caused by intersite-hybridization of the wave function and has to be calculated using the modern formulation~\cite{chang1996berry,xiao2005berry,thonhauser2005orbital,ceresoli2006orbital,gobel2018magnetoelectric} instead of relying on the often-used atomic-center approximation. So far, the effect has not been analyzed for more complicated structures, like CNTs, where the sub-band structure and Dirac physics make a classical comparison questionable.

Here we predict a chirality-induced selectivity of angular momentum in chiral CNTs. We consider a tight-binding model without spin-orbit coupling and use a Boltzmann approach to reveal that the orbital Edelstein susceptibility is purely governed by the chirality of the tube, as long as we consider metallic tubes near the Fermi level where the electrons are nearly mass-less Dirac fermions. This effect is nicely tunable by doping or application of a gate voltage making it highly significant for spinorbitronic applications. Further away from the Fermi level and for insulating tubes, the effect is strongly affected by the sub-band structure that can be understood by backfolding the band structure of a graphene lattice. \\
%
%
%
%
%
\\
\noindent\textbf{Results and Discussion}\\
\noindent\textbf{Carbon nanotubes}

\noindent Here we consider ideal CNTs that are periodic along the $z$ direction.
The unit cell can be constructed geometrically by rolling up a rectangular cut of a two-dimensional graphene layer. This layer is characterized by two basis atoms and the two lattice vectors $\vec{a}_1=a\vec{e}_x$ and $\vec{a}_2=(a/2)\vec{e}_x-(a\sqrt{3}/2)\vec{e}_y$ with the lattice constant $a=2.46\,$\AA. 

The rectangle that ultimately forms the unit cell of the nanotube is spanned by the circumferential vector $\vec{C}=n\vec{a}_1+m\vec{a}_2$ and the translational vector $\vec{T}=\frac{2m+n}{d}\vec{a}_1-\frac{2n+m}{d}\vec{a}_2$ with $d$ the greatest common divisor of $2m+n$ and $2n+m$. The values of $n\in \mathbb{N}^+$ and $m=0,1,\dots,n$ determine the configuration of the tube. If the tube was cut horizontally, a $(n,m)=(n,0)$ tube would have a zigzag edge and a $(n,m)=(n,n)$ tube would have an armchair edge. Both of these types of tubes are achiral but all other tubes, for which $m\neq0$ and $m\neq n$, are chiral. Note that tubes with the opposite chirality can be constructed by allowing also for negative values of $m$.

\begin{figure*}[th!]
    \centering
    \includegraphics[width=1\textwidth]{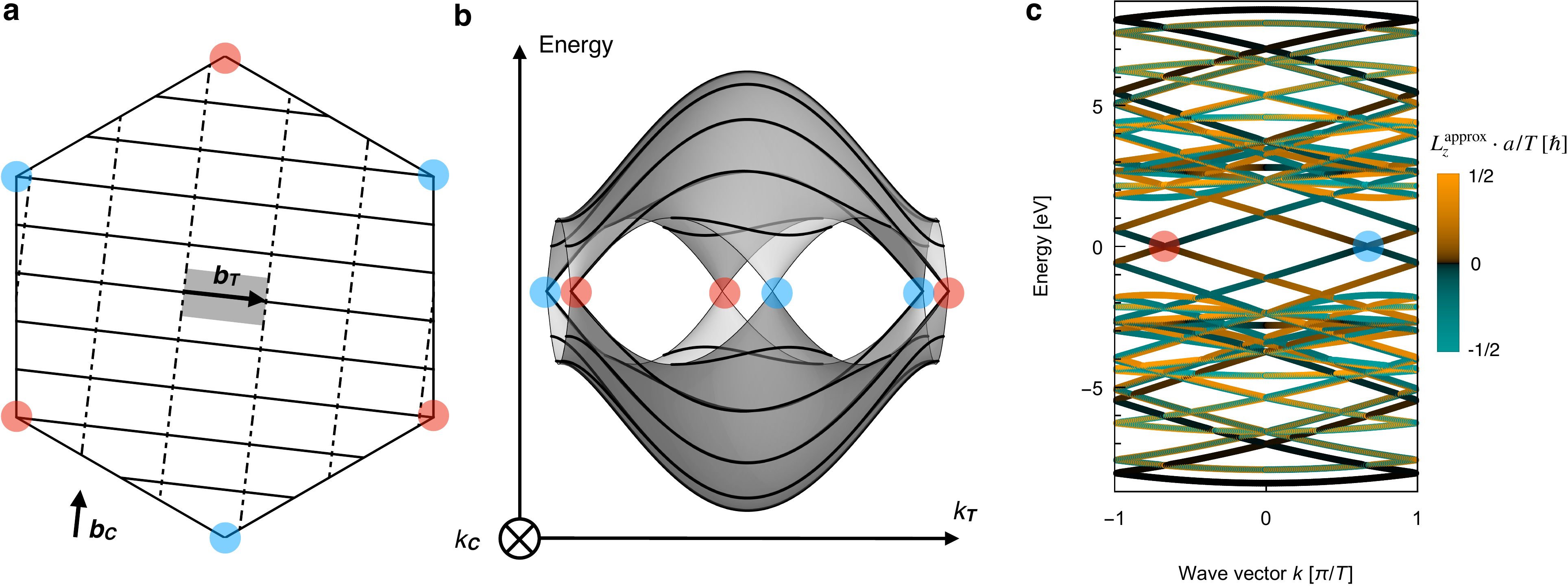}
    \caption{\textbf{Band structure of a metallic carbon nanotube by backfolding the graphene band structure.} \textbf{a} Brillouin zone of two-dimensional graphene (hexagon) and smaller Brillouin zone (gray) corresponding to the unit cell spanned by $\vec{C}$ and $\vec{T}$ (gray rectangle in Fig.~\ref{fig:nanotubes}a) for a $(n,m)=(7,1)$ tube. Black lines represent multiples of the reciprocal lattice vector $\vec{b}_T$ shifted by multiples of the reciprocal lattice vector $\vec{b}_C$. \textbf{b} Band structure of two-dimensional graphene. The black lines from panel \textbf{a} give rise to the band structure of the corresponding nanotube in \textbf{c} once they are shifted back to the small Brillouin zone (gray). The color resembles $L_{\nu,z}^\mathrm{approx}$ that approximates $L_{\nu,z}(k)$ of the corresponding carbon nanotube. The quantity has been normalized by $a/T$ for comparability. Colored dots indicate the Dirac points originally located at the points $K$ and $K'$ of the Brillouin zone of graphene.}
    \label{figS:backfolding_metal}
\end{figure*}

An example of the geometrical construction of a chiral CNT is shown in Fig.~\ref{fig:nanotubes}. The circumferential vector shown in \ref{fig:nanotubes}a is $\vec{C}=7\frac{1}{2}a\vec{e}_x-(a\sqrt{3}/2)\vec{e}_y=7\vec{a}_1+1\vec{a}_2$. The gray rectangle is rolled up to form the unit cell of the $(7,1)$ CNT in Fig.~\ref{fig:nanotubes}b. This nanotube is chiral with a chirality angle of
\begin{align}
    \alpha=\sphericalangle(\vec{T}_\mathrm{graphene},\vec{e}_y)=\arctan\frac{-m\sqrt{3}/2}{n+m/2},
\end{align}
in our example $\alpha\approx -6.59^\circ$. The angle can be recognized by following the color code in Fig.~\ref{fig:nanotubes}b along the translational direction $\vec{T}$. When we look into the tube from the top (cf. Fig.~\ref{fig:nanotubes}c), this color code results in a spiral as a signature of the chirality. Due to the 6-fold rotational symmetry of graphene, we will analyze only configurations with $|\alpha|\leq30^\circ$ and focus on the ones with negative chirality angles. Note, that the definition of the chirality angle $\alpha$ varies throughout the literature and can be defined with the opposite sign by choosing a different reference line of the graphene lattice. \\
%
%
%
%
%
%
\\
\noindent\textbf{Band structure and orbital angular momentum}\\
We use the geometrically constructed unit cell in a tight-binding model in which we consider the $p_z$ orbitals of the graphene layer. This single-orbital nature of the model describes the properties of a CNT well near the Fermi level and is characterized by an isotropic hopping amplitude $t$ and a vanishing spin-orbit coupling (for details see Methods).

Diagonalizing the tight-binding Hamiltonian results in the band structure and eigenvectors. The band structure is symmetric and exhibits many bands $E_{\nu}(k)$ due to the large unit cell. Since the tube is only periodic along the tube vector $\vec{T}\parallel\vec{e}_z$, we find dispersion only along $k\equiv k_z$. 
The system behaves quasi-one-dimensional. Note, however, that the chirality of the system is accounted for by the hopping paths along the $x$ and $y$ directions.

The band structure of our example nanotube $(n,m)=(7,1)$ (cf. Fig.~\ref{fig:bands_edelstein}a) consists of 76 bands and can be easily understood based on backfolding the graphene band structure~\cite{barros2006review,dresselhaus1998physical} ; cf. Fig.~\ref{figS:backfolding_metal}. Most importantly, the two Dirac points at the Fermi level of graphene exist also for this nanotube. This is the case for all metallic nanotubes, $\mathrm{mod}(n-m,3)=0$. All other nanotubes exhibit a band gap and have no Dirac points near the Fermi level and are characterized by $\mathrm{mod}(n-m,3)=1,2$; an example is shown in Fig.~\ref{fig:bands_edelstein}c.

In Fig.~\ref{fig:bands_edelstein}(a,c), the orbital angular momentum is added as a colorcode (normalized to a comparable length, $L_{\nu,z}(k)\cdot a/T$). It has been calculated based on the modern formulation of orbital magnetization (for details see Methods).
For the considered metallic nanotube, the values are in the order of up to $\sim\pm0.5\,\hbar$ and depend antisymmetrically on $k$. $L_{\nu,z}$ is exactly zero at the energy of the Dirac point. In fact, if we expand the system linearly around this energy, the orbital angular momentum is zero everywhere $L_{\nu,z}(k)\equiv0$. This can be interpreted as a consequence of the massless Dirac fermions. However, the actual band structure deviates from a perfect Dirac cone. As a consequence, $L_{\nu,z}(k)$ increases approximately linearly with $(k-k_\mathrm{Dirac})$. However, this dependence is slightly different for $k>k_\mathrm{Dirac}$ and $k<k_\mathrm{Dirac}$. For the other Dirac point, the orbital angular momentum is exactly reversed. This delicate $k$-dependence determines the orbital Edelstein effect in this energy range, as we will discuss later.

The orbital angular momentum is caused by the circular motion in the nanotube. \changeb{In equilibrium, it can even be determined analytically based on topological arguments according to Refs.~\cite{izumida2016angular,okuyama2019topological}.} Coming back to the two-dimensional graphene layer that forms the nanotube, this motion corresponds to a translational motion along the direction $\vec{C}$. Making use of the semiclassical formula $\vec{L}=\vec{r}\times\vec{p}$, we find for each band of the nanotube that $L_{\nu,z}$ is approximately given by the product of the radius of the tube $r=\nicefrac{C}{2\pi}$, the electron mass $m_e$ and the velocity $v_C=\frac{1}{\hbar}\frac{\partial E_\mathrm{2d}(k_T,k_C)}{\partial k_C}$ along the $\vec{C}$ direction which is calculated from the band structure $E_\mathrm{2d}(k_T,k_C)$ of the graphene layer as 
$L_{\nu,z}^\mathrm{approx}(k)\sim m_erv_C(k_T,k_C)=\frac{m_e}{\hbar}\frac{C}{2\pi}\frac{\partial E_\mathrm{2d}(k_T,k_C)}{\partial k_C}$.
Note that $k_T$ corresponds to $k$ in the nanotube and $k_C$ determines the sub-band index $\nu$. 

In summary, the symmetric band structure can be understood by backfolding the graphene band structure along the $k_T$ direction and the antisymmetric orbital angular momentum is proportional to the perpendicular velocity (along $k_C$) of graphene. While Fig.~\ref{fig:bands_edelstein}a shows these properties calculated directly from the CNT, Fig.~\ref{figS:backfolding_metal}c shows the results based on the backfolding, which are almost identical. A small difference occurs in $L_{\nu,z}(k)$ because the cross section of the CNT is not an actual cylinder but a polygon.\\
%
%
\\
\noindent\textbf{CIOS by orbital Edelstein effect}\\
\noindent The symmetric band structure $E_\nu(-k)=E_\nu(k)$ that exhibits an antisymmetric orbital angular momentum $L_{\nu,z}(-k)=-L_{\nu,z}(k)$ in reciprocal space (Fig.~\ref{fig:bands_edelstein}a) leads to the emergence of an orbital Edelstein effect. Since CNTs with opposite chirality have the same band structure but opposite orbital angular momentum, they give rise to exactly opposite orbital Edelstein susceptibilities (cf. blue and red curves in Fig.~\ref{fig:bands_edelstein}b). This means, CNTs give rise to a CIOS, i.\,e., they act as an orbital filter or orbital polarizer.

The orbital Edelstein susceptibility $\chi_z^{L_z}$ quantifies the non-equilibrium orbital magnetic moment per unit cell $m_{L_z}^{\mathrm{uc}}$ that is generated once an electric field $\boldsymbol{\mathcal{E}}=\mathcal{E}_z\,\vec{e}_z$ is applied,
$m_{L_z}^{\mathrm{uc}} = \chi_z^{L_z} \mathcal{E}_z$. For details about the calculation we refer to the Methods section.
Note that we will often discuss this quantity normalized to a unit tube length, $\chi_z^{L_z}\cdot a/T$, so that the signals of different tubes with different unit cells are comparable. 

In a quasi-one-dimensional system, the calculation of the orbital Edelstein susceptibility $\chi_z^{L_z}$ is especially easy to understand because it is simply proportional to the sum of all $L_{\nu,z}(k)\cdot\mathrm{sign}\{v_{\nu,z}(k)\}$ at each $k$ point corresponding to a given energy $E$. 
For energies close to the Dirac points, we find that this quantity increases quadratically with energy as a result of the compensation effect mentioned before. This makes the orbital Edelstein susceptibility highly tunable by doping or application of a gate voltage. 

The orbital Edelstein susceptibility per unit length of the tube $\chi_{z}^{L_z}\cdot a/T$ (Fig.~\ref{fig:bands_edelstein}b) is considerably large in that energy range with values up to $\sim 10^{-8}\mu_\mathrm{B}\,\mathrm{m}/\mathrm{V}$. At $E=0.49\,\mathrm{eV}$, (dashed line in Fig.~\ref{fig:bands_edelstein}) $\chi_{z}^{L_z}\cdot a/T\sim 10^{-9}\mu_\mathrm{B}\,\mathrm{m}/\mathrm{V}$. For energies farther away from the location of the Fermi level $E_\mathrm{F}=0$, we find oscillations with energy. Here, the orbital Edelstein susceptibility even reaches values of $\sim 10^{-7}\mu_\mathrm{B}\,\mathrm{m}/\mathrm{V}$. The oscillations can be understood based on the formation of the sub-band structure by backfolding, shown in Fig.~~\ref{figS:backfolding_metal}. The backfolding leads to different oscillations of the orbital Edelstein susceptibility for different metallic CNTs except for the energy region close to the Fermi level. Here, there are always only four bands that resemble the two Dirac points. The orbital Edelstein susceptibility for different metallic CNTs, including the (12,6) CNTs mentioned in the introduction, is shown in Fig.~S1a in the Supplementary Information and will be systematically analyzed below.

\begin{figure}[t!]
    \centering
    \includegraphics[width=\columnwidth]{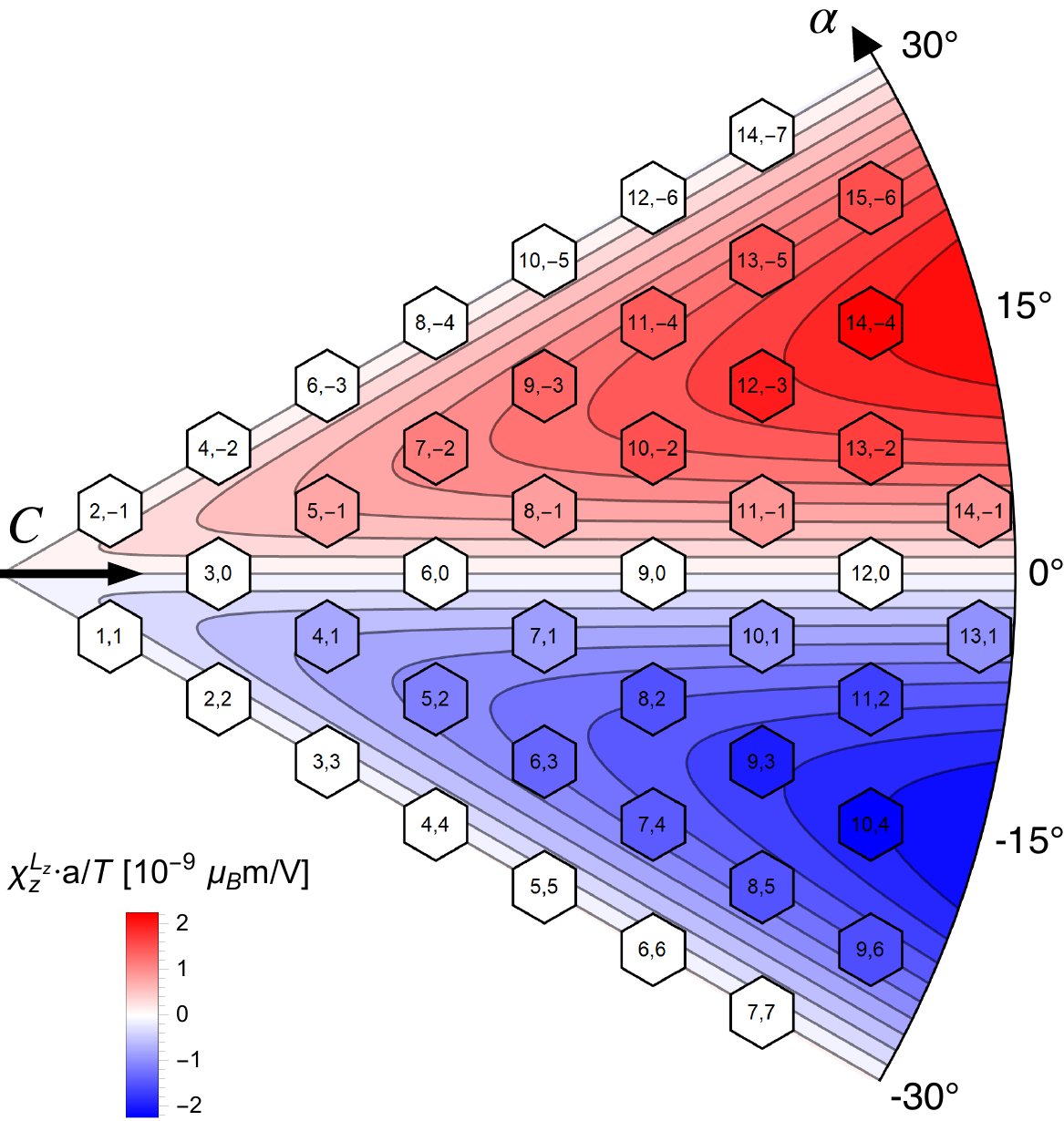}
    \caption{\textbf{Comparison of metallic carbon nanotubes.} The diagram shows all metallic nanotubes that are characterized by a circumference $C\leq14a$ (plotted along the radial direction) and a chirality angle $|\alpha|\leq30^\circ$ (plotted along the polar direction). The color of the hexagon indicates the value of the orbital Edelstein susceptibility at an energy of $E=0.49\,\mathrm{eV}$ (dashed line in Fig.~\ref{fig:bands_edelstein}). This value has been normalized by the length of the tube $T$, for comparability: $\chi_z^{L_z}\cdot a/T$ (see legend). The color in the background is the function proportional to $C\sin(6\alpha)$.}
    \label{fig:classification}
\end{figure}

Here we want to note that for the insulating CNTs, the backfolded band structure, shown in Fig.~S2 in the Supplementary Information, reveals that there are no Dirac points and that there is a band gap at the Fermi level. Still, the origin of the finite orbital Edelstein susceptibility is the same as for the metallic cases: We see a finite orbital angular momentum that is antisymmetric with $k$, as presented in Fig.~\ref{fig:bands_edelstein}c for a $(n,m)=(4,2)$ tube. Like for the metallic cases, there are oscillations in the energy-dependent orbital Edelstein susceptibility (Fig.~\ref{fig:bands_edelstein}d) that can be explained based on the backfolding. A comparison of various other insulating CNTs is shown in Fig.~S1b in the Supplementary Information.\\
%
%
%
\\
\noindent\textbf{Chirality dependence}\\
\noindent Even though the orbital Edelstein susceptibility of metallic CNTs is finite, it varies strongly with energy due to the sub-band structure that can be understood based on backfolding the graphene band structure. Yet, near the Fermi level, there are always only four bands that exhibit two Dirac points. This gives rise to a quadratic dependence of $\chi_z^{L_z}$ on the energy in all cases. This means, the behavior of all metallic CNTs is qualitatively the same in this energy range. 

\begin{figure*}[t!]
    \centering
    \includegraphics[width=0.95\textwidth]{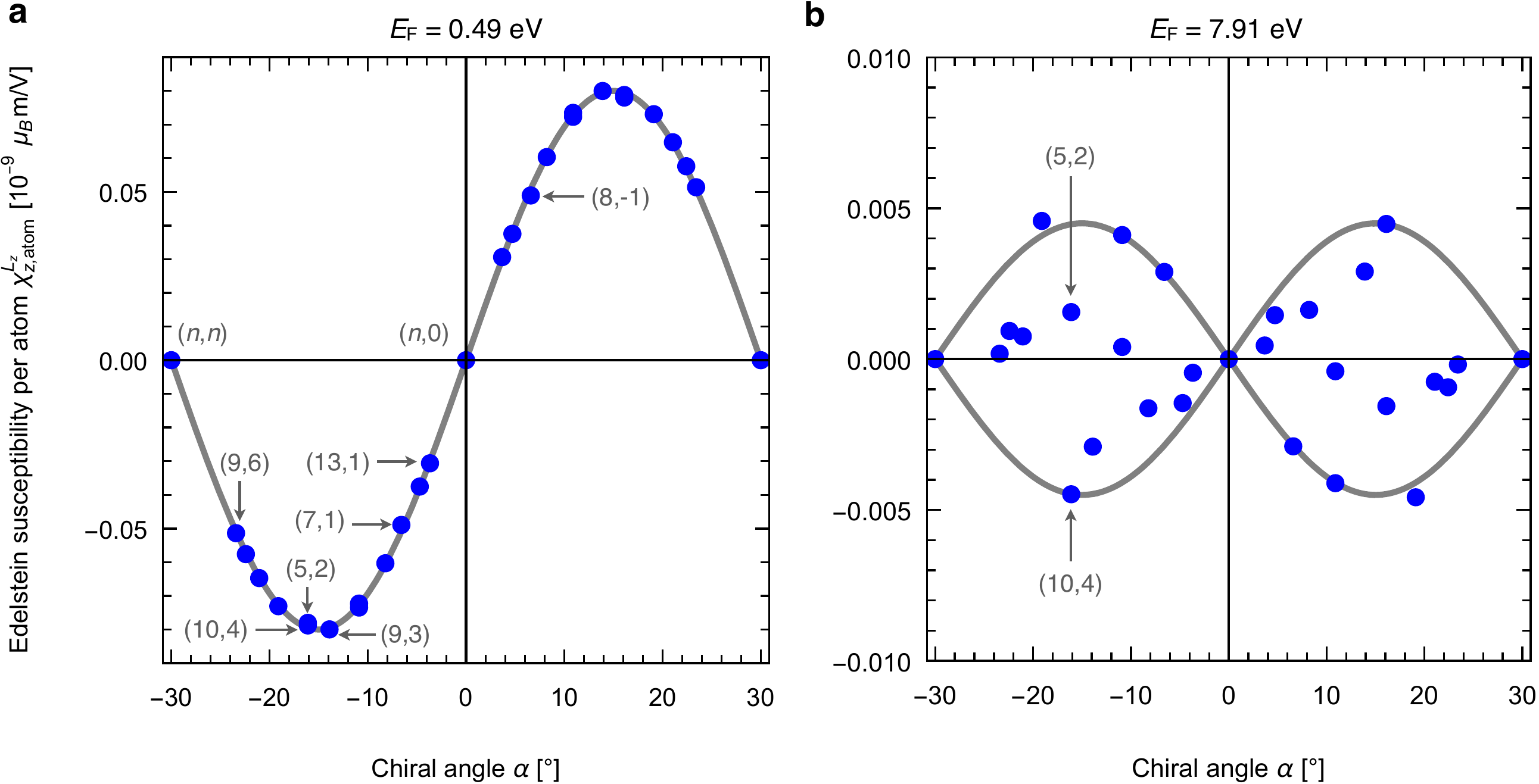}
    \caption{\textbf{Chirality dependence of the orbital Edelstein susceptibility per atom.} \textbf{a} The blue points are the calculated values $\chi_{z,\mathrm{atom}}^{L_z}$ at an energy of $E=0.49\,\mathrm{eV}$ for all metallic nanotubes shown in Fig.~\ref{fig:classification}. Here the band structure exhibits four nearly linear bands and the effect is governed by the chirality of the tube as can be seen by comparison with the gray line that is a function proportional to $\sin(6\alpha)$. \textbf{b} shows the equivalent plot at the energy $E=7.91\,\mathrm{eV}$ where the effect is mostly governed by the sub-band structure leading to pronounced differences of the orbital Edelstein susceptibility between the different nanotubes.}
    \label{fig:chirality}
\end{figure*}

In Fig.~\ref{fig:classification}, we systematically compare all metallic nanotubes with a circumference smaller than $14a\approx 3.4\,\mathrm{nm}$ (radius smaller than $\approx 0.55\,\mathrm{nm}$). In this diagram, each tube is resembled by a hexagon that is placed in a polar coordinate system: the distance from the origin is the tube's circumference $C$ and the polar angle is the chirality angle $\alpha$. The color of each hexagon resembles the normalized orbital Edelstein susceptibility $\chi_z^{L_z}\cdot a/T$ at an energy of $0.49\,\mathrm{eV}$ where the considered nanotubes exhibit always only 4 almost linear bands. The magnitude of $\chi_z^{L_z}\cdot a/T$ takes values of up to $\pm 2.2\times 10^{-9}\,\mu_\mathrm{B}\,\mathrm{m}/\mathrm{V}$ for the considered cases.

We find that $\chi_z^{L_z}\cdot a/T$ increases with the circumference of the tube $C$ and exhibits an antisymmetric dependence on the chirality angle $\alpha$. For the achiral cases, there is no orbital Edelstein effect: the zigzag tubes $(n,0)$ are characterized by $\alpha=0^\circ$ and the armchair tubes $(n,n)$ and $(2n,-n)$ by $\alpha=\pm 30^\circ$. The configurations that deviate the most from the two achiral configurations with a preserved inversion symmetry are chiral tubes characterized by $\alpha=\pm 15^\circ$. We see that $|\chi_z^{L_z}|\cdot a/T$ is largest near these angles. In the background of Fig.~\ref{fig:classification} we have added a contour plot that resembles a function proportional to $C\sin(6\alpha)$ that approximates the calculated values of $\chi_z^{L_z}\cdot a/T$ well. The function accounts for the 6-fold rotational symmetry of the graphene lattice.

To better understand the chirality dependence of the orbital Edelstein susceptiblity, we plot this quantity per atom $\chi_{z,\mathrm{atom}}^{L_z}=\chi_{z}^{L_z}\cdot\frac{1}{2}\cdot a^2\frac{\sqrt{3}}{2}/(C\cdot T)$ in Fig.~\ref{fig:chirality}a. We can nicely see a $\sin(6\alpha)$ dependence which indicates that the effect is purely governed by the structural chirality of the nanotube in this energy range. For example, both the $(5,2)$ and $(10,4)$ nanotubes are characterized by the same chirality angle $\alpha\approx-16.10^\circ$ and give rise to nearly the same $\chi_{z,\mathrm{atom}}^{L_z}$ (indicated by arrows). However, it is important to remember that $\chi_z^{L_z}\cdot a/T$ is the relevant quantity to compare different nanotubes and this quantity is twice as large for the $(10,4)$ nanotube compared to the $(5,2)$ nanotube due to the double circumference $C$. 

Lastly, we would like to note that this strict chirality dependence of the effect only is applicable in this energy range near the Fermi level. Fig.~\ref{fig:chirality}b shows the same plot at an energy of $E=7.91\,\mathrm{eV}$ where, depending on the tube, we see vastly different band structures and multiple states with different $L_{\nu,z}(k)$ exist. As a result, the blue data points are scattered and even the $(5,2)$ and $(10,4)$ nanotubes that are characterized by the same chirality angle have strongly different $\chi_{z,\mathrm{atom}}^{L_z}$, even with different signs. The values are $1.55\times 10^{-12}\,\mu_\mathrm{B}\,\mathrm{m}/\mathrm{V}$ versus $-4.48\times 10^{-12}\,\mu_\mathrm{B}\,\mathrm{m}/\mathrm{V}$, respectively.\\
\\
\noindent \change{\textbf{Relation to experiments and CISS}}\\
\noindent \change{In an experiment, the orbital magnetic moment generated by CIOS could be detected as a spin moment leading to CISS, due to a conversion in the electrodes caused by their spin-orbit coupling~\cite{liu2021chirality}. This means that the measured signal could be affected and even dominated by CIOS which is caused directly by the chirality of the sample and is not dependent on the often small spin-orbit coupling in a chiral structure.}

\change{However, note that a feature in transport experiments predicted by this approach requires further investigation~\cite{liu2021chirality}: When using magnetic leads, the $I$-$V$ dependence is predicted to become asymmetric due to the current-induced orbital magnetic moment. However, this asymmetry is not observed in most transport experiments, although an asymmetric signature is visible in Refs.~\cite{xie2011spin,bhowmick2022spin}. The orbital mechanism discussed in this paper is therefore likely not the only origin of CISS in chiral materials but is a complementary effect to the spin-driven effect that originates from spin-orbit coupling of the chiral material itself. From the theory side, it would be interesting to quantify corrections to the orbital Edelstein effect and to calculate the spin Edelstein effect caused by curvature-induced spin-orbit coupling~\cite{izumida2009spin,klinovaja2011carbon} and the hybridization of different orbitals. For large radii, we expect these corrections to be negligible. Experimentally, it would be helpful to vary the spin-orbit coupling of the lead (e.\,g. by changing the material) and to analyze the effect on the CISS signal in transport experiments, as previously suggested~\cite{evers2022theory}.}

When a magnetic field $B$ is applied along the CNT, \changeb{it couples to the orbital angular momentum} and affects the band structure accordingly~\cite{minot2004determination,kuemmeth2008coupling,marganska2005orbital}. In a unidirectional magnetoresistance measurement, the resistance $R$ depends on the sign of $B$ which gives rise to the phenomenon of electrical magnetochiral anisotropy (EMChA) that is a signature of chiral materials~\cite{rikken2001electrical,pop2014electrical,aoki2019anomalous}. We expect that this effect occurs for chiral CNTs as well but the precise field dependence and the interplay with the electric field that generates the non-equilibrium orbital magnetic moment has to be analyzed carefully in future studies.

\change{Besides the aforementioned potential for spin-to-charge interconversion, chiral CNTs could also be used directly for orbital-to-charge interconversion. The experimental observation of orbital magnetic moments and orbital currents has progressed rapidly over the past few years, as summarized in the review paper~\cite{jo2024spintronics}. Detection has been achieved: (i) By magneto-optical Kerr effect measurements~\cite{choi2023observation,lyalin2023magneto}, (ii) by detecting charge currents corresponding to inverse orbital effects~\cite{el2023observation}, (iii) by measuring the Hanle magnetoresistance~\cite{sala2023orbital}, (iv) via measuring orbital torques~\cite{lee2021orbital}. Especially noteworthy for our findings is a previous measurement of the inverse orbital Edelstein effect in a two-dimensional electron gas at an oxide interface~\cite{el2023observation}. Based on our calculations of the orbital Edelstein susceptibility, chiral CNTs act as filters and polarizers for orbital currents which could potentially be detected by the mentioned techniques and mechanisms.} 

\change{Furthermore,  CIOS could be measured more easily by photoemission than in transport experiments~\cite{liu2021chirality}: Chiral molecules play the same role as spiral phase plates for photoelectrons and their total detectable magnetic moment includes the dominating orbital angular momentum. This means that the spin moment, that may come from the spin-orbit coupling of the sample or substrate, would not interfere with the orbital contribution as much as in transport experiments.}\\
%
\\
\noindent\textbf{Conclusion}\\
\noindent In summary, we have shown that CNTs exhibit a chirality-dependent selectivity of orbital angular momentum. The orbital Edelstein susceptibility is exactly opposite for two chiral nanotubes with the opposite chirality and otherwise the same properties. The effect does not occur for the achiral zigzag and armchair nanotubes. While the sub-band structure strongly influences the energy dependence of the effect, all metallic nanotubes exhibit only 4 nearly linear bands near the Fermi level. Here, the orbital Edelstein susceptibility depends quadratically on the energy; the effect can be nicely tuned by applying a gate voltage or by doping. 

Near the Fermi level, the effect is purely governed by the chirality of the structure and scales linearly with the circumference of the nanotube. However, note that this means that the orbital Edelstein susceptibility per cross-sectional area $A$ scales inversely with the circumference, $m_{L_z}/A\propto\frac{1}{C}\sin(6\alpha)\mathcal{E}_z$. For this reason, growing many small nanotubes in a finite area, can result in a larger CIOS than having only a few large nanotubes, as long as the nanotubes can be positioned closely together. It might also be interesting to consider multi-walled CNTs where multiple nanotubes can be positioned inside of each other, thereby increasing or decreasing the effect, depending on the respective chiralities.

Finding an approximate relation between the orbital angular momentum $L_z$ in CNTs and the velocity $v_C$ in graphene along $\vec{C}$ (as presented before) is also an important result because it allows us to connect our findings with previous calculations from the 1990s. In fact, before the modern formulation of orbital magnetization was established~\cite{xiao2005berry,thonhauser2005orbital} and before CISS has been observed~\cite{ray1999asymmetric}, a `chiral conductivity' had been theoretically discussed that is calculated via $v_C$ of the two-dimensional material forming the tube (for BC$_2$N nanotubes~\cite{miyamoto1996chiral} and for BN nanotubes~\cite{kral2000photogalvanic}). This chiral conductivity has even been related to self-inductance~\cite{miyamoto1999self} meaning that a chiral current gives rise to magnetic properties. Therefore, the chiral conductivity is closely related to the orbital Edelstein susceptibility that we calculate here giving rise to CIOS. Indeed, similar oscillating energy-dependent curves of the chiral conductivity have been calculated~\cite{lambert2008oscillating} but the authors only considered insulating CNTs.

Our results show that the orbital angular momentum couples to the chirality of a system and exists even without spin-orbit coupling. In semi-classical terms, the origin of CIOS is in real-space as there are helical hopping paths along chiral CNTs. However, in comparison to CIOS in a simple helix~\cite{gobel2025chirality}, as in tellurium or selenium, the Dirac physics and sub-band structure in CNTs have an important influence on the effect. They make a quantum mechanical treatment crucial, while in a toy model of a helix~\cite{gobel2025chirality} the effect was perfectly analogous to a classical motion along a helix. In comparison to a conventional Rashba system, where an Edelstein effect occurs only at an interface with the magnetization perpendicular to the current~\cite{aronov1989nuclear,edelstein1990spin}, here we generate a homogeneous magnetization all over the CNT along the current direction.

In future studies, related chiral systems like nanotubes formed by multiple elements, chiral nano-carbons, chiral carbon-based heterostructures or molecular-functionalized nanotubes can be analyzed, for some of which CISS has been observed experimentally already~\cite{alam2015spin,alam2017spin,rahman2018long,rahman2020carrier,rahman2021molecular,hao2021direct,rahman2022chirality}. The efficient and controllable filtering of spin and orbital angular momentum allows for the generation of spin and orbital torques. This makes chiral CNTs promising candidates for applications in quantum computing and spin-orbitronics.\\
\\
\\
\\
\\
\textbf{Methods}\\
\textbf{Hamiltonian}\\
\noindent The tight-binding Hamiltonian \change{considers the radial $p_z$ orbitals of the carbon atoms}
\begin{align}
    H=-t\,\sum_{\braket{i,j}} c_i^\dagger c_j.
\end{align}
\change{Other orbitals are not required to resemble ab initio calculations, especially for large tube radii~\cite{correa2010tight}.} Here, $c_i^\dagger$ and $c_j$ are the creation and annihilation operators of an electron at site $i$ and $j$, respectively. The bracket indicates nearest neighbors. In our example of a $(n,m)=(7,1)$ nanotube, the unit cell consists of $76$ atoms, each with 3 nearest neighbors. $t=2.8\,\mathrm{eV}$ is the hopping amplitude. 

\change{We consider periodic copies of the tube along $x$ and $y$ but do not consider any coupling between them. Therefore, even though the band structure exhibits no dispersion along $k_x$ and $k_y$, the Hamiltonian depends on these inplane components of the wave vector which accounts for the chirality of the nanotube. The individual hopping terms from atom $j$ to $i$ have a phase $\vec{k}\cdot\vec{r}_{ij}$. Here $\vec{r}_{ij}$ is the connecting vector that has in general $x$, $y$ and $z$ components.} \changeb{By using this convention, the eigenvectors correspond to the lattice-periodic part of the Bloch wave from which the orbital angular momentum can be calculated directly.}\\
\\
\textbf{Orbital angular momentum}\\
\noindent We calculate the $z$ component of the orbital angular momentum according to the modern formulation of orbital magnetization~\cite{chang1996berry,xiao2005berry,thonhauser2005orbital,ceresoli2006orbital,gobel2018magnetoelectric}
\begin{align}
      &L_{\nu,z}(k) = \mathrm{i} \frac{m_e}{g_L\hbar} \sum_{\mu \neq \nu} \frac{1}{E_{\mu}(k) - E_{\nu}(k)}\\
      &\times\left[\braopket{\nu k}{\nicefrac{\partial H}{\partial k_x}}{\mu k} \braopket{\mu k}{\nicefrac{\partial H}{\partial k_y}}{\nu k} - (\nu\leftrightarrow \mu)\right].\notag
\end{align}
Here, $\nu$ is the band index with $\ket{\nu k}$ the eigenvector and $E_\nu(k)$ the eigenenergy determined by diagonalizing the Hamiltonian $H$. \changeb{Note that the eigenvectors correspond to the lattice-periodic part of the Bloch wave.} $m_e$ is the electron mass, $g_L=1$ is the Landé g-factor and $\hbar$ the reduced Planck constant. Note that the trick of using the super cell is equivalent to defining $L_{\nu,z}$ by means of a real-space tight-binding Hamiltonian. This formulation includes inter-site contributions that are missing in the atomic-center approximation, which is often used in the literature.\\
\\
\textbf{Orbital Edelstein susceptibility}

\noindent The orbital Edelstein susceptibility $\chi_z^{L_z}$ is calculated using the Boltzmann transport theory
\begin{align}
    \chi_{z}^{L_z}(E)&=\frac{e^2g_L}{2m_e}\times\\
    &\sum_{\nu,k} \tau_{\nu}(k)\cdot v_{\nu,z}(k)\cdot L_{\nu,z}(k)\cdot\delta(E_\nu(k)-E)\notag
\end{align}
with $v_{\nu,z}(k)=\frac{1}{\hbar}\frac{\partial E_\nu(k)}{\partial k}$ the one-dimensional group velocity along the tube. Only states with the energy $E$ contribute, due to the delta distribution $\delta$. We use a constant relaxation time $\tau_{\nu}(k)\equiv\tau_0=1\,\mathrm{ps}$. Note that this is an approximation and that the values of $\chi_{z}^{L_z}$, discussed in the main text, scale with the assumed $\tau_0$. The quantitative values might therefore not be fully reliable but the results are significant qualitatively.\\
\\
\textbf{Data availability}\\
The data that support the findings of this work are available at DOI: 10.5281/zenodo.17065253.\\
\\
\textbf{Code availability}\\
The code that supports the findings of this work is available from the authors on reasonable request.\\
\\
\textbf{Acknowledgements}\\
This work was supported by the EIC Pathfinder OPEN grant 101129641 ``Orbital Engineering for Innovative Electronics''.\\
\\
\textbf{Author contributions}\\
B.G. performed calculations, prepared the figures, planned the project and wrote the manuscript with significant inputs from I.M. and S.L.. B.G., I.M. and S.L. discussed the results.\\
\\
\textbf{Additional information}\\
\textbf{Supplementary information} The online version contains supplementary material available at [insert link].\\
\\
\textbf{Competing interests}\\
The authors declare no competing interests.\\
\\
\textbf{References}




\end{document}